\documentclass[aps,prb,twocolumn,floatfix,superscriptaddress]{revtex4-1}

\usepackage{graphicx}
\usepackage{amsmath}
\usepackage{amssymb}
\usepackage{braket}
\usepackage{nicematrix}
\usepackage[normalem]{ulem}

\newcommand{\bea}{\begin{eqnarray}}
	\newcommand{\eea}{\end{eqnarray}}
\newcommand{\beq}{\begin{equation}}
	\newcommand{\eeq}{\end{equation}}

\begin{document}

\title{Microtubules as electron-based topological insulators}
\author{Varsha Subramanyan}
\email{varshas2@illinois.edu}
\affiliation{Dept. of Physics, University of Illinois, Urbana-Champaign,USA}
\author{Kay L. Kirkpatrick}
\affiliation{Dept. of Mathematics, University of Illinois, Urbana-Champaign,USA}
\affiliation{Dept. of Physics, University of Illinois, Urbana-Champaign,USA}
\author{Saraswathi Vishveshwara}
\affiliation{Molecular Biophysics Unit, Indian Institute of Science, Bengaluru, India}
\author{Smitha Vishveshwara}
\email{smivish@illinois.edu}
\affiliation{Dept. of Physics, University of Illinois, Urbana-Champaign,USA}


\begin{abstract}
The microtubule is a cylindrical biological polymer that plays key roles in cellular structure, transport, and signalling.  In this work, based on studies of electronic properties of polyacetelene and mechanical properties of microtubules  themselves\cite{prodan}, we explore the possibility that microtubules could act as topological insulators that are gapped to electronic excitations in the bulk but possess robust electronic bounds states at the tube ends. Through analyses of structural and electronic properties, we  model the microtubule as a cylindrical stack of Su-Schrieffer-Heeger chains (originally proposed in the context of polyacetylene) describing electron hopping between the underlying dimerized tubulin lattice sites. We postulate that the microtubule is mostly uniform, dominated purely by GDP-bound dimers, and is capped by a disordered regime due to the presence of GTP-bound dimers as well. In the uniform region, we identify the electron hopping parameter regime in which the microtubule is a topological insulator. We then show the manner in which these topological features remain robust when the hopping parameters are disordered. We briefly mention possible biological implications for these microtubules to possess topologically robust electronic bound states. 
\end{abstract}
\maketitle

Organic polymers and biomolecules have served as inspiration for discovering new condensed matter phenomena, with path-breaking insights developed for both biological systems and unique physical models\cite{Spakowitz,Calvo}. The Su-Schrieffer-Heeger (SSH) model proposed in the 1980's for polyacetylene serves even today as the simple paradigm for topological phases of matter and associated charge fractionalization\cite{SSH}. Such models have then led to opening entire fields of study, bringing theoretical and experimental advances in our understanding of materials\cite{Asboth}. On the biological side, the fundamental units of the model, involving dimerization and sublattice symmetry, are the building blocks for a range of complex macromolecules. Here, we investigate microtubules as such a highly promising instance for extending topological behavior in rich and diverse ways. 

The microtubule is a crucial protein complex that forms an integral part of the cellular cytoskeleton and consists of dimerized units arranged in a helical structure\cite{Wade2009,Amos2011}. In Ref. \onlinecite{prodan},  the authors target the mechanical properties of these dimer units embedded in a manifold in the uniform (non-disordered) case and model the underlying interactions as harmonic coupling. The resulting dynamical equations show similarities with the Hamiltonian of a topological insulator.  Strikingly, the similarity translates to the mechanical system being topological in that its phases can be characterized by certain invariants and topologically non-trivial regimes in parameter space possess a class of solutions reflecting signature end modes rendered robust by a spectral gap. As with metamaterials\cite{XIN2020100853}, the authors associate these modes with robust boundary phonon modes that play an important role in the growth and decay phases of microtubules. We do the same in a corresponding electronic system, motivated by the healthy interplay between metamaterial and electronic physics. The crucial difference is that here the systems are one and the same--we expect that the predicted phonon behavior in microtubules would imply corresponding analogues in its electronic properties. 

\begin{figure}
    \centering
    \includegraphics[width=0.5\textwidth]{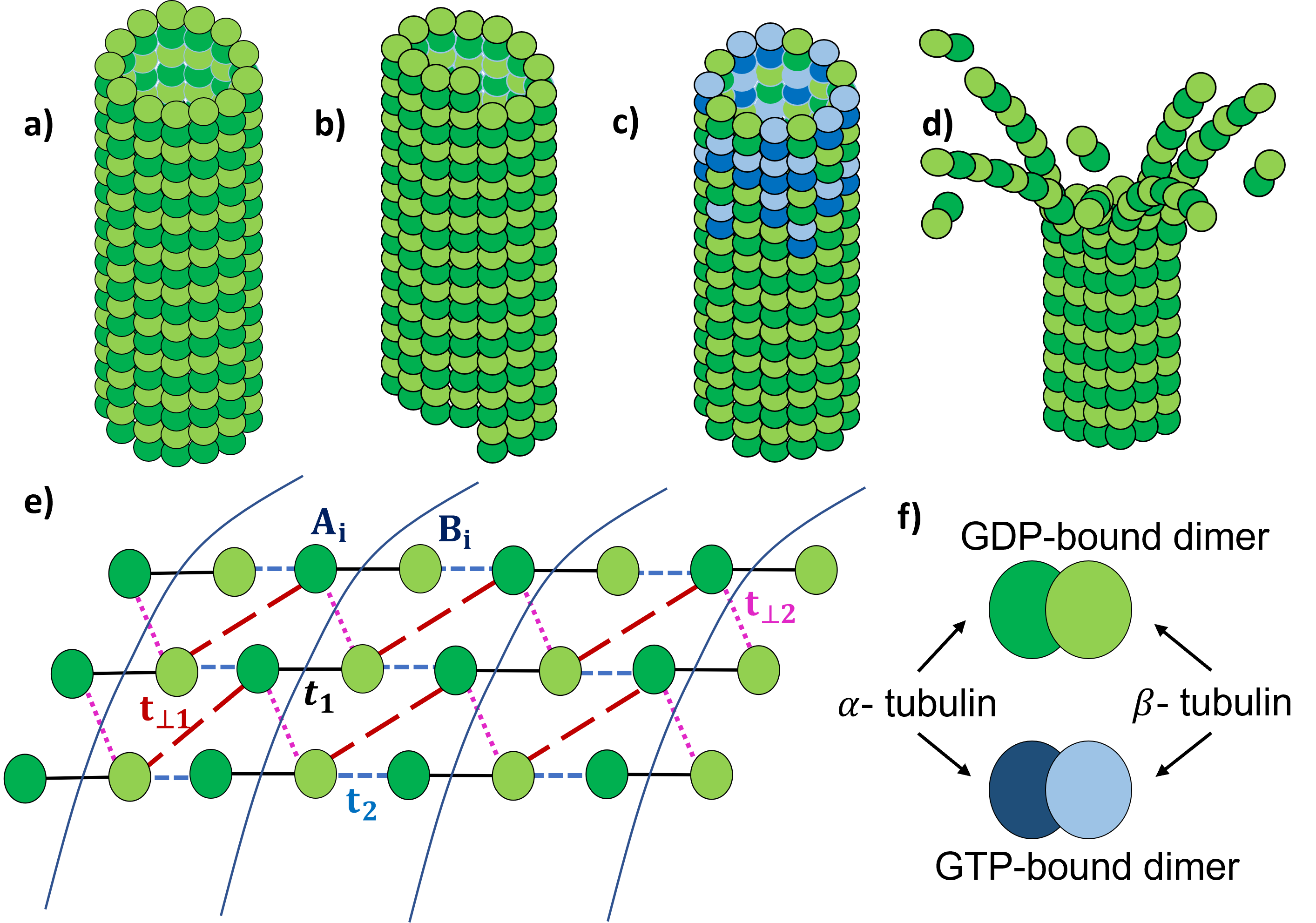}
    \caption{The cylindrical microtubule is shown here in different phases and configurations. It is composed mainly of GDP-bound dimers (shades of green). Figure a) shows a microtubule of zero pitch, while Figure b) shows a microtubule having pitch equal to one dimer's length. Figures c) and d) show the microtubule in two possible phases. While a) and b) are uniform, in c), the ``cap" of the microtubule has dimers of both the GDP and GTP (blue) kind, leading to increase in its length when more dimers from the surrounding medium attach onto the cap. This non-uniformity prevents the peeling off or reduction in microtubule length, depicted in Figure d). Figure e) shows the lattice model of the microtubule with associated hopping strengths indicated. In Figure f) the $\alpha$ and $\beta$ tubulin monomers are indicated in each type of dimer. }
    \label{hop}
\end{figure}
The attribution of topological properties to electronic systems has important consequences \cite{Bernevig2013}. For instance, fractional charge polarization, originally analyzed in polyacetelene \cite{Meier2016}and a zero bias peak in the differential conductivity are defining signatures of topological phases. These quantities are also robust to small amounts of disorder permeating the system due to the presence of a finite energy gap. However, it is also known that stronger disorder strengths mediate phase transitions between topological and trivial phases.

In this work, we show that within a regime of plausible parameter windows, microtubules can possess a topological phase determined by their electronic band structure, alongside the mechanical phase posited in Ref. \onlinecite{prodan}. On consideration of electronic processes inherent to the microtubule, we model it as a cylindrical stack of SSH chains. We explore the properties of such quasi one-dimensional weak topological insulators and their response to disorder. We then discuss how these electronic phases map back to biological properties, possibly showing radically different kinds of conduction properties of microtubules stemming from the presence of topologically robust bound states.

{\bf Biophysics, structure, and electronic properties: } Microtubules play important roles in  maintaining structural integrity of the cell, material transport within it, in initiating cellular mitosis, and in facilitating intracellular signalling\cite{Avila1992-ko}. They are prominent structures in eukaryotic cells, especially crucial to the morphological make-up of neurons. The microtubule has long been a popular candidate for the exploration of quantum mechanical effects manifesting in biological systems\cite{HAMEROFF1982549,HAMEROFF1996453}. It has often been argued that finite temperature and the environment associated with living cells would lead to decoherence and suppression of such effects. However, several recent studies provide evidence that such effects may not be ignored in several biological systems including microtubules\cite{Marais,Craddock,Mavromatos}. Here, we base our fundamental assumptions on such evidence.

Structurally, the microtubule is a helical cylinder composed of tubulin molecules as the fundamental unit. These units are dimers consisting the homologous monomers $\alpha-$tubulin and $\beta-$tubulin. Typically, the cylinders have a 25 nm outer diameter and 15 nm inner diameter. The dimers themselves have dimensions of 8 nm $\times$ 5 nm $\times$ 5 nm\cite{Tuszynski2018}.  The dimers are attached helically around the cylinder giving rise to protofilaments, usually 13 in number around the circumference. The cylindrical lattice thus formed is highly ordered and is most commonly seen in two different configurations - with and without a prominent seam, corresponding to differences in helical pitch, as seen in Fig 1. As has been done with regards to mechanical properties in Ref. \onlinecite{prodan}, these cylinders can thus be modeled as extensions of the SSH system where the chain is wrapped around in a cylindrical configuration. 

The microtubule shows two distinct phases when it undergoes ``dynamic instability" - a growth or polymerization phase, where dimers from the surrounding solution attach themselves to the filaments of the microtubule, and a ``peeling off" or de-polymerization or catastrophe phase, where the filaments of the microtubule curve outward longitudinally and lose dimers to the solution, decreasing the effective length of the tube\cite{Brouhard2015,Horio1}. The microtubule has been observed to stochastically switch between the two phases. The dimers themselves can be of two possible types distinguished by the molecule attached at an exchange site - GDP or GTP. During the growth phase, dimers of both kinds attach themselves to the microtubule to form longer filaments. This non-uniform segment at the end of tube is called the cap. As the tube grows, the GTP hydrolyzes to GDP after attachment to form uniform GDP attached extended regions. The GDP-attached filaments are longitudinally curved, and it has been proposed that this curvature that results in the ``peeling" off when the energetics overcome the stability provided by the cap\cite{Burbank2006-yd,Brun21173}. Based on this structure of the microtubule, in our model, we distinguish two distinct regimes---one in the uniform GDP-based region, which we model as consisting of no disorder, and the other in the cap region, which we model as having a disorder distribution in the inter-tubulin charge hopping amplitudes.



Turning to the electronic properties, ample computational and observational studies indicate that the dimerized structure of the microtubule affect not only vibrational degrees of freedom, but also the intersite electronic hopping within the underlying lattice\cite{JT1,Tuszynski2019,Minoura2006}.  
The conduction electron concentration on a microtubule has been estimated to be $n \sim 10^{19}cm^{-3}$, which is higher than that of semiconductors, but significantly lower than that of metals. Even lower estimates of its conductivity from these numbers (at physiological temperatures) suggest a conductivity\cite{Tuszynski2018} that lies between $0.04 \Omega^{-1}m^{-1}$ and $10^5\Omega^{-1}m^{-1}$. 
The microtubule has a large net negative charge per tubulin dimer that is mostly concentrated on the ''C-terminus'' of each dimer. The uncompensated negative charge is balanced in solution by counterions that screen the charge, giving rise to a dielectric polarization around the surface of the microtubule.  This observation has led to several models of conductivity in terms of tight-binding of electrons to the lattice site and effective charge hopping from site to site, like in the case of organic semiconductors\cite{Ruhle2011,Kirkpatrick2013}. Additionally, since the surrounding medium of the microtubule is rich in positively charged ions\cite{Tuszynski2019,Eakins}, the highly negatively charged microtubule might also exhibit conduction by proton hopping mechanisms, leading to higher conductivity, a phenomenon that plays a secondary role in this work. These descriptions for conductivity form the starting point of our analysis; we model electronic properties in terms of electrons hopping between nearest neighbor lattice sites provided by the underlying $\alpha$- and $\beta$-tubulin units. 



{\bf Effective Quasi-1D model: }Having established this tight-binding description for electron hopping between the $\alpha$ and $\beta$ sites of the microtubule, we first analyze the uniform region. The resultant cylindrical lattice and hopping strengths are as depicted in Fig. \ref{hop}. The system consists of SSH chains along the length of the tube having alternating bond strengths $t_1$ and $t_2$. As shown in the figure, along the circumferential direction, the units are connected via bonds $t_{\perp, 1}$ and $t_{\perp, 2}$. Depending on the pitch of the helix, there will be conformational changes in the helix, which can be reflected in the relative strength between $t_{\perp, 1}$ and $t_{\perp, 2}$. The pitch will also be reflected in the allowed quantized values of $k_y$. Here, due to the helical nature of the tubule, we assume that the staggering of the chains results in these inter-chain hoppings dominating. Thus, the predominant hopping terms respect sub-lattice symmetry in that hoppings are only between sub-lattices $A$ consisting of $\alpha$ sites and $B$ consisting of $\beta$ sites, but not within each sublattice. The resultant Hamiltonian thus takes the generic form:
\begin{align}
     H&=\sum_{ij}\begin{pmatrix}
    c_{A,ij}^\dagger & c_{B,ij}^\dagger
    \end{pmatrix}
    \mathbb{H}
    \begin{pmatrix}
    c_{A,ij} \\ c_{B,ij}
    \end{pmatrix}
\end{align}

where $\mathbb{H}$ obeys chiral symmetry such that $S\mathbb{H}S=-\mathbb{H}$ with $ S=\sum_{ij} (c^\dagger_{A,ij}c_{A,ij}-c^\dagger_{B,ij}c_{B,ij})$. This Hamiltonian, emerging from the dominant hoppings, respects the symmetries of  quasi one dimensional topological systems belonging to class AIII, classified as per the usual Altland-Zirnbauer formalism that characterizes topological systems according to their symmetries\cite{Schnyder}. 

\begin{figure}
    \centering
    \includegraphics[width=0.5\textwidth]{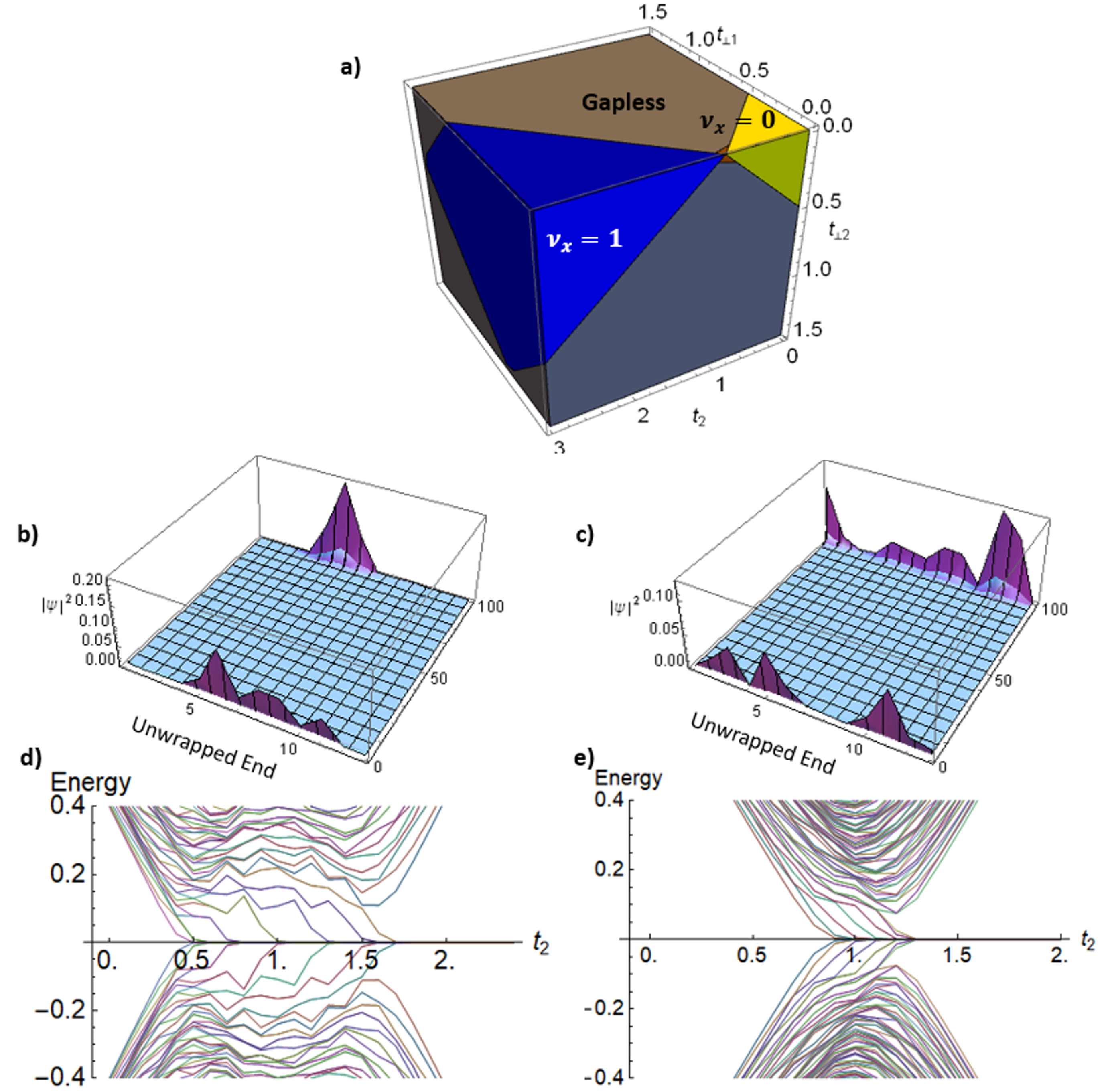}
    \caption{Energy bands and phase diagram: Figure a) shows the phase diagram for the lattice model under the thermodynamic limit, as a function of hopping strengths $t_2$, $t_{\perp,1}$ and $t_{\perp,2}$ all scaled to be in units of $t_1$.The gapless and gapped phases are indicated, with the gapped phases being identified by their associated topological invariant. Figures b) and c) show one of the possible zero energy edge states for a cylindrical strip of $N_y=13$ chains with 50 dimers along the length of the microtubule in the topological phase. Figure b) is for a helical cylinder of pitch zero, while c) is for a helical cylinder with unit pitch.  Figures d) and e) demonstrate the changes in the band structure for strip of pitch zero (Figure d)) and pitch one (Figure e)) when the parameter $t_2$ is tuned from 0 to 3. Here, $t_{\perp,1}=0.2$ and $t_{\perp,2}=0.1$. The figure shows an initial configuration with a trivial gap become gapless for a range of values before again opening up a topological gap. The zero energy modes are also seen in the topological gap. }
    \label{band}
\end{figure}

A qualitative picture of how an 'end-mode' can form along the circumference at the two ends of a segment of a uniform region is as follows.  In the absence of interchain hopping, the paradigm SSH chains running along the length of the tube naturally support end bound states. These bound states correspond to robust mid-gap states in a spectrum symmetric about the Fermi energy having a gap of magnitude $|t_1-t_2|$. Upon inter-chain coupling, these states too couple to form the end-mode along the rim\cite{Delplace,Zhu}; the robustness of similar states has been discussed in the context of coupled ladder systems \cite{Li, Hugel,Padavic}.  Figure.\ref{band} b) and c) plot such typical end-mode eigenstates in a regime that we have determined to be topological. For strong enough interchain coupling, the gap in the spectrum closes and the system undergoes a transition into a phase where no such robust end-mode exists. 



To analyze these features more rigorously, we express the Hamiltonian for a segment of this lattice in momentum space. To first order, we assume that the segment is long enough to take the thermodynamic limit. We assume a unit lattice constant for each bond. 
The resultant Hamiltonian takes the form  

\begin{align}
    H&=\begin{pmatrix}
    c_{A,\vec{k}}^\dagger & c_{B,\vec{k}}^\dagger
    \end{pmatrix}
    \begin{pmatrix}
    0&q(\vec{k})\\q^\dagger(\vec{k})&0
    \end{pmatrix}
    \begin{pmatrix}
    c_{A,\vec{k}} \\ c_{B,\vec{k}}
    \end{pmatrix}
    \label{H}
\end{align}

where $q(\bar k)=u+ve^{-ik_x}$, $\bar k=(k_x,k_y)$,  $u=t_1+2t_{\perp, 2}\cos{k_y}$ and $v=2t_{\perp, 1}\cos{ky}+t_2$. The energy band structure is given by
\begin{align}
     E_\pm&=\pm \sqrt{u^2+v^2+2uv\cos{k_x}}.
\end{align}
In order to chart out the regimes where the system is topological in that a finite-sized tube would have end modes present, we can use the standard topological analysis of the band structure. In the thermodynamic limit, or under periodic boundary conditions in both directions, the band structure is gapped for a range of parameter regimes, as shown in the phase diagram in Fig. \ref{band}. The gapped phases may further be classified as topological or trivial, as depicted in the phase diagram. Crucially, the topological phase is characterized by a topological invariant or winding number defined as 
\begin{align}
    \nu_x=-\frac{i}{2\pi}\int dk_x Tr[q^{-1}\partial_{k_x}q]\label{nu}.
\end{align}
In this system, $\nu$ takes the value 1 for $|u|<|v| $ (topological) or 0  for $|u|>|v|$ (trivial)\cite{Asboth}. The system is gapless when $|u|=|v|$.
 
The band structure of a finite cylindrical strip is shown in Fig. \ref{band} d) and e), for two different boundary conditions of the microtubule. In the figures, as parameter $t_2$ is increased, it is seen that that system initially has a trivial gap, and then passes through a regime of gaplessness before opening up a topological gap. The topological phase is distinguished by the presence of zero energy (compared to the Fermi energy) edge states.  For a cylindrical strip of $N_y$ chains, there are $2N_y$ zero energy modes in the gap, in the absence of sub-lattice symmetry breaking terms. In our quasi 1D model, this situation corresponds to a topological invariant $\nu_x=1$ (higher integers are usually associated with long-range coupling in such conjugated systems). These edge states can then couple with other systems, hybridize with each other for short enough tube lengths, or interact with environmental modes. These effects give rise to a host of phenomena that have been extensively studied in the context of Hermitian as well as non-Hermitian SSH chains and would be of direct relevance to microtubules\cite{Lieu,McGinley,Campos}. Moreover, while the staggered helical nature of the microtubule allows the reasonable assumption of low hoppings of the kind A-A or B-B or non-uniform on-site chemical potential, their presence explicitly breaks the aforementioned sublattice symmetry of the system. Its effects may be understood in terms of the effects of extended nest-nearest-neighbour coupling in SSH chains\cite{NNN}. For small values of such symmetry breaking terms (with respect to the gap), the spectrum is shifted or stretched or both, thus lifting some or all the edge states from zero energy. However, they retain their topological significance despite due to the presence of the gap. Very high values, on the other hand, can close the gap and cause these edge states to become absent.


{\bf Disorder and Dynamic Instability: }While the above analysis is most relevant to the uniform regime corresponding to the peeling-off segment, by the cap theory, the growth phase of the microtubule sets in when the ends are not uniformly composed of GDP dimers, but also contain interspersed GTP dimers. Further, in some diseased states, there could also be isoforms of tubulin present, which would further contribute to the non-uniformity of the cap\cite{Vemu2017-zh}. This non-uniformity can be reasonably expected to affect the electron hopping strength. Moreover, given that the GTP molecules are in metastable stages of hydrolysis enroute to conversion to GDP, we expect a continuous range of non-uniformity, as opposed to a binary distribution.  We model the non-uniformity in the cap as disorder in the hopping parameters, and characterize its effects on topological properties via changes made to the winding number. 

Since the cap does not cover the entire length of the microtubule, we assume disorder only in part of the tube. As a specific proof-of-principle instance, we model the presence of disorder in half the tube. In order to determine whether or not the tube is in the topological phase, we can no longer employ the momentum basis form for deriving the invariant of Eq.\ref{nu} since disorder breaks translational invariance. We may however employ the real space version typically used for disordered systems \cite{EProdan,Claes,Hughes}. To briefly provide the recipe, in the basis of eigenstates of $S$, the Hamiltonian can be written in the off-diagonal form of Eq.\ref{H}. One then defines a 'flat-band' version $Q$ of this real-space Hamiltonian as
$Q=P_+-P_-$, where $P_\pm$ are projectors onto eigenspaces with positive and negative eigenvalues. 
The invariant is given by
\begin{align}
    \nu=-\frac{1}{N_x}Tr(\Tilde{Q}^\dagger[X,\Tilde{Q}]), \textnormal{ }Q=\begin{pmatrix}
    0&\Tilde{Q}\\\Tilde{Q}^\dagger&0
    \end{pmatrix}.
\end{align}
Here $N_x$ is the number of unit cells along the lattice and $X$ is the position operator in the lattice. 

\begin{figure}
   \centering
   \includegraphics[width=0.5\textwidth]{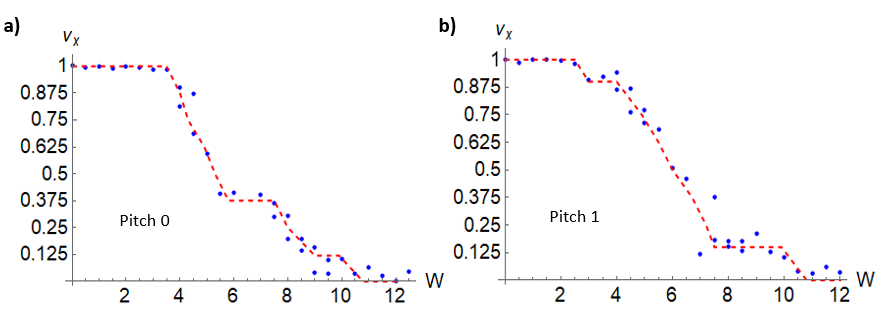}
    \caption{Topological invariant $\nu_x$ plotting against disorder strength W in $t_{1}$ and strength W/2 in $t_{2}$ for $N_y$=8 in two possible pitch configurations. The blue points correspond to the real space topological invariant for a microtubule of pitch zero, and the red curve is a rough fit of the data points. The initial quantization is briefly retained until $W\sim 3.5$. The system briefly stabilizes at $\nu_x=3/8$ before the gap closes for stronger disorder.}
    \label{dis}
\end{figure}

Our model of the microtubule corresponds to the AIII class of topological insulators. 
We effectively evaluate the topological invariant per unit width $\nu_x=\nu/N_y$ where $N_y$ is the number of protofilaments. Thus, $\nu_x$ is quantized in units of $1/N_y$. This quantized behaviour has been established both analytically and numerically, even in the presence of strong disorder\cite{Claes}. In our system, we apply disorder to the hopping strengths $t_1$ and $t_2$ individually. For each of these hopping strengths (indicated generally as $h_i$), disorder is applied through the equation $h_i=h_{i0}+W_iu$ where $u$ is a uniform random variable sampled between $[-0.5,0.5]$ and $W_i$ is the strength of the disorder. The initial parameters $h_{i0}$ are picked so that the microtubule starts at a topological phase. We also pick $W_1=W$ for $t_1$ and $W_2=W/2$ for $t_2$. These parameters represent a typical sample which demonstrates the plausible evolution of topological features.

The results thus obtained are shown in Fig.\ref{dis}. A few caveats are in order.The trend clearly shows a nearly quantized value of unity for the invariant at low disorder that drops towards zero upon increasing disorder. The curve can stabilize at other values of $\nu_x$ that are multiples of $1/N_y$. Fluctuations around the allowed values are a result of the finite size of the system chosen and they diminish for longer systems\cite{Claes}.  Given the computational power involved, our strips were confined to  $N_y=8$. We expect the results to hold for the more realistic number $N_y\sim 13$. While the form of the invariant assumes sublattice symmetry, we expect the topological feature of a robust edge mode to remain for small deviations from this symmetry condition. Most importantly, the simulation is a demonstration of the salient properties in the presence of disorder. Our simulations show that the quantization is retained for values of disorder as high as $W=3.5$.
 While this treatment considers the segment as a whole, with disorder in half the region to represent the physical situation, a more detailed analysis would show the exact physical location of the end modes as possibly shifting into the interior. Most prominently, we may reasonably expect that for regions of lower disorder, the presence of the edge mode would still be prevalent. 

{\bf Discussion and Outlook: }Here, we have given a proof-of-concept presentation of how topological properties of tight-binding lattices can provide biological microtubules with unique electronic bound state structure hitherto unexplored.  In particular, we associate the decay phase of the microtubule with the presence of an electronic topological edge mode, while we associate the growth phase with the disordered case that could possibly lead to the vanishing of the edge mode. Several variations of course would come into play with regards to the rudimentary model presented here. To mention a few, in this work, we assumed sublattice symmetry, which ultimately gives rise to the mid-gap states centered at the Fermi energy. We expect that deviations either due to on-site potential variations or other couplings within each of the $A$ and $B$ sublattices would still preserve the edge modes for small enough values. On the other hand, for large enough deviations, these edge states would disappear; perhaps the asymmetry of the physical properties in the two ends of the tubes could emerge from related differences in the chemical composition.  Additionally, one could adapt these calculations to extremely short microtubules as they start to grow out of the centrosome, and perhaps to extremely long microtubules that push an embryonic neuronal cell membrane out to grow towards future synapses. Two other important factors to consider are coupling to the environment and the non-equilibrium nature of tube growth and dynamic instability. 

With regards to disorder, while our model is phenomenological in nature, further studies would connect the disorder distribution to physical conditions, such as  hydrolysis states from GTP to GDP, temperature effects, and presence of other biochemicals. Moreover, here, while we studied uniform versus disordered behavior for entire segments, it is possible to have more robust bound states at the interface of such segments. It is postulated that when isotypes of tubulin are present in diseased microtubules, their functioning is only marginally altered, with such defects only becoming apparent in processes that are highly sensitive to microtubule dynamics\cite{Gadadhar}. This view additionally supports the presence of topologically robust modes in the microtubule. Another important consideration is the previous study on mechanical degrees of freedom demonstrating that the uniform region is unstable due to a topological phonon mode, in contrast to the electronic modes studied here. We expect both modes to follow similar trends in the presence of disorder. A further intriguing possibility would entail electron-phonon coupling. 

Finally, the existence of robust edge bound state modes could have significant biological implications. In principle, the bound state could act as an acceptor or donor of charge, perhaps differentially influencing the more negatively charged GTP compared with the GDP molecule for the processes discussed here. The bound state could also be affected by microtubule associated proteins, such as as kinesin and dynein.  Microtubules are key to intracellular long-range transport, for instance, quite strikingly in neural transmission. In addition, if edge modes of microtubules provide a localized state for electronic occupation, then they may extend into various parts of the axon terminal, and dynamically change the open-or-closed states of voltage-gated ion channels, contributing to pre-synaptic membrane activity. Conversely, proteins that have an overall positive charge or neurite regions that have a concentration of positive charge (such as an action potential's influx of cations) might attract the C-terminus of a microtubule if it is in the topological regime for hosting negative charge at its end. Regardless of the extent to which microtubules are involved with intra-neuronal activity besides transport \cite{HAMEROFF1982549,HAMEROFF1996453}, it is likely that sub-neuronal computation is an important part of brain function, and microtubules may play a significant role in that.
As in condensed matter systems, conduction properties of microtubules hinge greatly on their structure, chemical composition and environment; their possible behavior as topological insulators may bear radical insights into the quantum nature of biophysical systems. 

{\bf Acknowledgements: }We thank Camelia Prodan and Emil Prodan for sharing beautiful insights both in their work and through conversations. We also thank Kesav Krishnan for useful conversations. We acknowledge the support of the National Science Foundation under grants DMR-2004825 (V. S. and Sm.V.) and CAREER DMS-1254791 (K. L. K.).  Sa.V. is an Honorary Scientist of the National Academy of Sciences, India(NASI).

\bibliography{references}

\begin{thebibliography}{42}%
\makeatletter
\providecommand \@ifxundefined [1]{%
 \@ifx{#1\undefined}
}%
\providecommand \@ifnum [1]{%
 \ifnum #1\expandafter \@firstoftwo
 \else \expandafter \@secondoftwo
 \fi
}%
\providecommand \@ifx [1]{%
 \ifx #1\expandafter \@firstoftwo
 \else \expandafter \@secondoftwo
 \fi
}%
\providecommand \natexlab [1]{#1}%
\providecommand \enquote  [1]{``#1''}%
\providecommand \bibnamefont  [1]{#1}%
\providecommand \bibfnamefont [1]{#1}%
\providecommand \citenamefont [1]{#1}%
\providecommand \href@noop [0]{\@secondoftwo}%
\providecommand \href [0]{\begingroup \@sanitize@url \@href}%
\providecommand \@href[1]{\@@startlink{#1}\@@href}%
\providecommand \@@href[1]{\endgroup#1\@@endlink}%
\providecommand \@sanitize@url [0]{\catcode `\\12\catcode `\$12\catcode
  `\&12\catcode `\#12\catcode `\^12\catcode `\_12\catcode `\%12\relax}%
\providecommand \@@startlink[1]{}%
\providecommand \@@endlink[0]{}%
\providecommand \url  [0]{\begingroup\@sanitize@url \@url }%
\providecommand \@url [1]{\endgroup\@href {#1}{\urlprefix }}%
\providecommand \urlprefix  [0]{URL }%
\providecommand \Eprint [0]{\href }%
\providecommand \doibase [0]{http://dx.doi.org/}%
\providecommand \selectlanguage [0]{\@gobble}%
\providecommand \bibinfo  [0]{\@secondoftwo}%
\providecommand \bibfield  [0]{\@secondoftwo}%
\providecommand \translation [1]{[#1]}%
\providecommand \BibitemOpen [0]{}%
\providecommand \bibitemStop [0]{}%
\providecommand \bibitemNoStop [0]{.\EOS\space}%
\providecommand \EOS [0]{\spacefactor3000\relax}%
\providecommand \BibitemShut  [1]{\csname bibitem#1\endcsname}%
\let\auto@bib@innerbib\@empty
\bibitem [{\citenamefont {Prodan}\ and\ \citenamefont {Prodan}(2009)}]{prodan}%
  \BibitemOpen
  \bibfield  {author} {\bibinfo {author} {\bibfnamefont {E.}~\bibnamefont
  {Prodan}}\ and\ \bibinfo {author} {\bibfnamefont {C.}~\bibnamefont
  {Prodan}},\ }\href {\doibase 10.1103/PhysRevLett.103.248101} {\bibfield
  {journal} {\bibinfo  {journal} {Phys. Rev. Lett.}\ }\textbf {\bibinfo
  {volume} {103}},\ \bibinfo {pages} {248101} (\bibinfo {year}
  {2009})}\BibitemShut {NoStop}%
\bibitem [{\citenamefont {Spakowitz}(2019)}]{Spakowitz}%
  \BibitemOpen
  \bibfield  {author} {\bibinfo {author} {\bibfnamefont {A.~J.}\ \bibnamefont
  {Spakowitz}},\ }\href {\doibase 10.1063/1.5126852} {\bibfield  {journal}
  {\bibinfo  {journal} {The Journal of Chemical Physics}\ }\textbf {\bibinfo
  {volume} {151}},\ \bibinfo {pages} {230902} (\bibinfo {year} {2019})},\
  \Eprint {http://arxiv.org/abs/https://doi.org/10.1063/1.5126852}
  {https://doi.org/10.1063/1.5126852} \BibitemShut {NoStop}%
\bibitem [{\citenamefont {ALVAREZ-ESTRADA}\ and\ \citenamefont
  {CALVO}(2002)}]{Calvo}%
  \BibitemOpen
  \bibfield  {author} {\bibinfo {author} {\bibfnamefont {R.~F.}\ \bibnamefont
  {ALVAREZ-ESTRADA}}\ and\ \bibinfo {author} {\bibfnamefont {G.~F.}\
  \bibnamefont {CALVO}},\ }\href {\doibase 10.1080/00268970210121623}
  {\bibfield  {journal} {\bibinfo  {journal} {Molecular Physics}\ }\textbf
  {\bibinfo {volume} {100}},\ \bibinfo {pages} {2957} (\bibinfo {year}
  {2002})},\ \Eprint
  {http://arxiv.org/abs/https://doi.org/10.1080/00268970210121623}
  {https://doi.org/10.1080/00268970210121623} \BibitemShut {NoStop}%
\bibitem [{\citenamefont {Su}\ \emph {et~al.}(1980)\citenamefont {Su},
  \citenamefont {Schrieffer},\ and\ \citenamefont {Heeger}}]{SSH}%
  \BibitemOpen
  \bibfield  {author} {\bibinfo {author} {\bibfnamefont {W.~P.}\ \bibnamefont
  {Su}}, \bibinfo {author} {\bibfnamefont {J.~R.}\ \bibnamefont {Schrieffer}},
  \ and\ \bibinfo {author} {\bibfnamefont {A.~J.}\ \bibnamefont {Heeger}},\
  }\href {\doibase 10.1103/PhysRevB.22.2099} {\bibfield  {journal} {\bibinfo
  {journal} {Phys. Rev. B}\ }\textbf {\bibinfo {volume} {22}},\ \bibinfo
  {pages} {2099} (\bibinfo {year} {1980})}\BibitemShut {NoStop}%
\bibitem [{\citenamefont {Asbóth}\ \emph {et~al.}(2016)\citenamefont
  {Asbóth}, \citenamefont {Oroszlány},\ and\ \citenamefont
  {Pályi}}]{Asboth}%
  \BibitemOpen
  \bibfield  {author} {\bibinfo {author} {\bibfnamefont {J.~K.}\ \bibnamefont
  {Asbóth}}, \bibinfo {author} {\bibfnamefont {L.}~\bibnamefont {Oroszlány}},
  \ and\ \bibinfo {author} {\bibfnamefont {A.}~\bibnamefont {Pályi}},\ }\href
  {\doibase 10.1007/978-3-319-25607-8} {\bibfield  {journal} {\bibinfo
  {journal} {Lecture Notes in Physics}\ } (\bibinfo {year} {2016}),\
  10.1007/978-3-319-25607-8}\BibitemShut {NoStop}%
\bibitem [{\citenamefont {Wade}(2009)}]{Wade2009}%
  \BibitemOpen
  \bibfield  {author} {\bibinfo {author} {\bibfnamefont {R.~H.}\ \bibnamefont
  {Wade}},\ }\href {\doibase 10.1007/s12033-009-9193-5} {\bibfield  {journal}
  {\bibinfo  {journal} {Molecular Biotechnology}\ }\textbf {\bibinfo {volume}
  {43}},\ \bibinfo {pages} {177} (\bibinfo {year} {2009})}\BibitemShut
  {NoStop}%
\bibitem [{\citenamefont {Amos}(2011)}]{Amos2011}%
  \BibitemOpen
  \bibfield  {author} {\bibinfo {author} {\bibfnamefont {L.~A.}\ \bibnamefont
  {Amos}},\ }\href@noop {} {\bibfield  {journal} {\bibinfo  {journal} {Semin
  Cell Dev Biol}\ }\textbf {\bibinfo {volume} {22}},\ \bibinfo {pages} {916}
  (\bibinfo {year} {2011})}\BibitemShut {NoStop}%
\bibitem [{\citenamefont {Xin}\ \emph {et~al.}(2020)\citenamefont {Xin},
  \citenamefont {Siyuan}, \citenamefont {Harry}, \citenamefont {Minghui},\ and\
  \citenamefont {Yanfeng}}]{XIN2020100853}%
  \BibitemOpen
  \bibfield  {author} {\bibinfo {author} {\bibfnamefont {L.}~\bibnamefont
  {Xin}}, \bibinfo {author} {\bibfnamefont {Y.}~\bibnamefont {Siyuan}},
  \bibinfo {author} {\bibfnamefont {L.}~\bibnamefont {Harry}}, \bibinfo
  {author} {\bibfnamefont {L.}~\bibnamefont {Minghui}}, \ and\ \bibinfo
  {author} {\bibfnamefont {C.}~\bibnamefont {Yanfeng}},\ }\href {\doibase
  https://doi.org/10.1016/j.cossms.2020.100853} {\bibfield  {journal} {\bibinfo
   {journal} {Current Opinion in Solid State and Materials Science}\ }\textbf
  {\bibinfo {volume} {24}},\ \bibinfo {pages} {100853} (\bibinfo {year}
  {2020})}\BibitemShut {NoStop}%
\bibitem [{\citenamefont {Bernevig}(2013)}]{Bernevig2013}%
  \BibitemOpen
  \bibfield  {author} {\bibinfo {author} {\bibfnamefont {B.~A.}\ \bibnamefont
  {Bernevig}},\ }\href {\doibase doi:10.1515/9781400846733} {\emph {\bibinfo
  {title} {Topological Insulators and Topological Superconductors}}}\ (\bibinfo
   {publisher} {Princeton University Press},\ \bibinfo {year}
  {2013})\BibitemShut {NoStop}%
\bibitem [{\citenamefont {Meier}\ \emph {et~al.}(2016)\citenamefont {Meier},
  \citenamefont {An},\ and\ \citenamefont {Gadway}}]{Meier2016}%
  \BibitemOpen
  \bibfield  {author} {\bibinfo {author} {\bibfnamefont {E.~J.}\ \bibnamefont
  {Meier}}, \bibinfo {author} {\bibfnamefont {F.~A.}\ \bibnamefont {An}}, \
  and\ \bibinfo {author} {\bibfnamefont {B.}~\bibnamefont {Gadway}},\ }\href
  {\doibase 10.1038/ncomms13986} {\bibfield  {journal} {\bibinfo  {journal}
  {Nature Communications}\ }\textbf {\bibinfo {volume} {7}},\ \bibinfo {pages}
  {13986} (\bibinfo {year} {2016})}\BibitemShut {NoStop}%
\bibitem [{\citenamefont {Avila}(1992)}]{Avila1992-ko}%
  \BibitemOpen
  \bibfield  {author} {\bibinfo {author} {\bibfnamefont {J.}~\bibnamefont
  {Avila}},\ }\href@noop {} {\bibfield  {journal} {\bibinfo  {journal} {Life
  Sci}\ }\textbf {\bibinfo {volume} {50}},\ \bibinfo {pages} {327} (\bibinfo
  {year} {1992})}\BibitemShut {NoStop}%
\bibitem [{\citenamefont {Hameroff}\ and\ \citenamefont
  {Watt}(1982)}]{HAMEROFF1982549}%
  \BibitemOpen
  \bibfield  {author} {\bibinfo {author} {\bibfnamefont {S.~R.}\ \bibnamefont
  {Hameroff}}\ and\ \bibinfo {author} {\bibfnamefont {R.~C.}\ \bibnamefont
  {Watt}},\ }\href {\doibase https://doi.org/10.1016/0022-5193(82)90137-0}
  {\bibfield  {journal} {\bibinfo  {journal} {Journal of Theoretical Biology}\
  }\textbf {\bibinfo {volume} {98}},\ \bibinfo {pages} {549} (\bibinfo {year}
  {1982})}\BibitemShut {NoStop}%
\bibitem [{\citenamefont {Hameroff}\ and\ \citenamefont
  {Penrose}(1996)}]{HAMEROFF1996453}%
  \BibitemOpen
  \bibfield  {author} {\bibinfo {author} {\bibfnamefont {S.}~\bibnamefont
  {Hameroff}}\ and\ \bibinfo {author} {\bibfnamefont {R.}~\bibnamefont
  {Penrose}},\ }\href {\doibase https://doi.org/10.1016/0378-4754(96)80476-9}
  {\bibfield  {journal} {\bibinfo  {journal} {Mathematics and Computers in
  Simulation}\ }\textbf {\bibinfo {volume} {40}},\ \bibinfo {pages} {453}
  (\bibinfo {year} {1996})}\BibitemShut {NoStop}%
\bibitem [{\citenamefont {Marais}\ \emph {et~al.}(2018)\citenamefont {Marais},
  \citenamefont {Adams}, \citenamefont {Ringsmuth}, \citenamefont {Ferretti},
  \citenamefont {Gruber}, \citenamefont {Hendrikx}, \citenamefont {Schuld},
  \citenamefont {Smith}, \citenamefont {Sinayskiy}, \citenamefont {Krüger},
  \citenamefont {Petruccione},\ and\ \citenamefont {van Grondelle}}]{Marais}%
  \BibitemOpen
  \bibfield  {author} {\bibinfo {author} {\bibfnamefont {A.}~\bibnamefont
  {Marais}}, \bibinfo {author} {\bibfnamefont {B.}~\bibnamefont {Adams}},
  \bibinfo {author} {\bibfnamefont {A.~K.}\ \bibnamefont {Ringsmuth}}, \bibinfo
  {author} {\bibfnamefont {M.}~\bibnamefont {Ferretti}}, \bibinfo {author}
  {\bibfnamefont {J.~M.}\ \bibnamefont {Gruber}}, \bibinfo {author}
  {\bibfnamefont {R.}~\bibnamefont {Hendrikx}}, \bibinfo {author}
  {\bibfnamefont {M.}~\bibnamefont {Schuld}}, \bibinfo {author} {\bibfnamefont
  {S.~L.}\ \bibnamefont {Smith}}, \bibinfo {author} {\bibfnamefont
  {I.}~\bibnamefont {Sinayskiy}}, \bibinfo {author} {\bibfnamefont {T.~P.~J.}\
  \bibnamefont {Krüger}}, \bibinfo {author} {\bibfnamefont {F.}~\bibnamefont
  {Petruccione}}, \ and\ \bibinfo {author} {\bibfnamefont {R.}~\bibnamefont
  {van Grondelle}},\ }\href {\doibase 10.1098/rsif.2018.0640} {\bibfield
  {journal} {\bibinfo  {journal} {Journal of The Royal Society Interface}\
  }\textbf {\bibinfo {volume} {15}},\ \bibinfo {pages} {20180640} (\bibinfo
  {year} {2018})}\BibitemShut {NoStop}%
\bibitem [{\citenamefont {Craddock}\ \emph {et~al.}(2014)\citenamefont
  {Craddock}, \citenamefont {Friesen}, \citenamefont {Mane}, \citenamefont
  {Hameroff},\ and\ \citenamefont {Tuszynski}}]{Craddock}%
  \BibitemOpen
  \bibfield  {author} {\bibinfo {author} {\bibfnamefont {T.~J.~A.}\
  \bibnamefont {Craddock}}, \bibinfo {author} {\bibfnamefont {D.}~\bibnamefont
  {Friesen}}, \bibinfo {author} {\bibfnamefont {J.}~\bibnamefont {Mane}},
  \bibinfo {author} {\bibfnamefont {S.}~\bibnamefont {Hameroff}}, \ and\
  \bibinfo {author} {\bibfnamefont {J.~A.}\ \bibnamefont {Tuszynski}},\ }\href
  {\doibase 10.1098/rsif.2014.0677} {\bibfield  {journal} {\bibinfo  {journal}
  {Journal of The Royal Society Interface}\ }\textbf {\bibinfo {volume} {11}},\
  \bibinfo {pages} {20140677} (\bibinfo {year} {2014})}\BibitemShut {NoStop}%
\bibitem [{\citenamefont {Mavromatos}\ and\ \citenamefont
  {Nanopoulos}(1998)}]{Mavromatos}%
  \BibitemOpen
  \bibfield  {author} {\bibinfo {author} {\bibfnamefont {N.~E.}\ \bibnamefont
  {Mavromatos}}\ and\ \bibinfo {author} {\bibfnamefont {D.~V.}\ \bibnamefont
  {Nanopoulos}},\ }\href {\doibase 10.1142/S0217979298000326} {\bibfield
  {journal} {\bibinfo  {journal} {International Journal of Modern Physics B}\
  }\textbf {\bibinfo {volume} {12}},\ \bibinfo {pages} {517} (\bibinfo {year}
  {1998})},\ \Eprint
  {http://arxiv.org/abs/https://doi.org/10.1142/S0217979298000326}
  {https://doi.org/10.1142/S0217979298000326} \BibitemShut {NoStop}%
\bibitem [{\citenamefont {Tuszynski}\ \emph {et~al.}(2018)\citenamefont
  {Tuszynski}, \citenamefont {Priel}, \citenamefont {Brown}, \citenamefont
  {Cantiello},\ and\ \citenamefont {Dixon}}]{Tuszynski2018}%
  \BibitemOpen
  \bibfield  {author} {\bibinfo {author} {\bibfnamefont {J.~A.}\ \bibnamefont
  {Tuszynski}}, \bibinfo {author} {\bibfnamefont {A.}~\bibnamefont {Priel}},
  \bibinfo {author} {\bibfnamefont {J.~A.}\ \bibnamefont {Brown}}, \bibinfo
  {author} {\bibfnamefont {H.~F.}\ \bibnamefont {Cantiello}}, \ and\ \bibinfo
  {author} {\bibfnamefont {J.~M.}\ \bibnamefont {Dixon}},\ }\href@noop {}
  {\bibfield  {journal} {\bibinfo  {journal} {Nano and Molecular Electronics
  Handbook}\ } (\bibinfo {year} {2018})}\BibitemShut {NoStop}%
\bibitem [{\citenamefont {Brouhard}(2015)}]{Brouhard2015}%
  \BibitemOpen
  \bibfield  {author} {\bibinfo {author} {\bibfnamefont {G.~J.}\ \bibnamefont
  {Brouhard}},\ }\href {\doibase 10.1091/mbc.E13-10-0594} {\bibfield  {journal}
  {\bibinfo  {journal} {Molecular biology of the cell}\ }\textbf {\bibinfo
  {volume} {26}},\ \bibinfo {pages} {1207} (\bibinfo {year} {2015})},\ \bibinfo
  {note} {25823928[pmid]}\BibitemShut {NoStop}%
\bibitem [{\citenamefont {Horio}\ and\ \citenamefont {Murata}(2014)}]{Horio1}%
  \BibitemOpen
  \bibfield  {author} {\bibinfo {author} {\bibfnamefont {T.}~\bibnamefont
  {Horio}}\ and\ \bibinfo {author} {\bibfnamefont {T.}~\bibnamefont {Murata}},\
  }\href {\doibase 10.3389/fpls.2014.00511} {\bibfield  {journal} {\bibinfo
  {journal} {Frontiers in Plant Science}\ }\textbf {\bibinfo {volume} {5}},\
  \bibinfo {pages} {511} (\bibinfo {year} {2014})}\BibitemShut {NoStop}%
\bibitem [{\citenamefont {Burbank}\ and\ \citenamefont
  {Mitchison}(2006)}]{Burbank2006-yd}%
  \BibitemOpen
  \bibfield  {author} {\bibinfo {author} {\bibfnamefont {K.~S.}\ \bibnamefont
  {Burbank}}\ and\ \bibinfo {author} {\bibfnamefont {T.~J.}\ \bibnamefont
  {Mitchison}},\ }\href@noop {} {\bibfield  {journal} {\bibinfo  {journal}
  {Curr Biol}\ }\textbf {\bibinfo {volume} {16}},\ \bibinfo {pages} {R516}
  (\bibinfo {year} {2006})}\BibitemShut {NoStop}%
\bibitem [{\citenamefont {Brun}\ \emph {et~al.}(2009)\citenamefont {Brun},
  \citenamefont {Rupp}, \citenamefont {Ward},\ and\ \citenamefont
  {N{\'e}d{\'e}lec}}]{Brun21173}%
  \BibitemOpen
  \bibfield  {author} {\bibinfo {author} {\bibfnamefont {L.}~\bibnamefont
  {Brun}}, \bibinfo {author} {\bibfnamefont {B.}~\bibnamefont {Rupp}}, \bibinfo
  {author} {\bibfnamefont {J.~J.}\ \bibnamefont {Ward}}, \ and\ \bibinfo
  {author} {\bibfnamefont {F.}~\bibnamefont {N{\'e}d{\'e}lec}},\ }\href
  {\doibase 10.1073/pnas.0910774106} {\bibfield  {journal} {\bibinfo  {journal}
  {Proceedings of the National Academy of Sciences}\ }\textbf {\bibinfo
  {volume} {106}},\ \bibinfo {pages} {21173} (\bibinfo {year} {2009})},\
  \Eprint
  {http://arxiv.org/abs/https://www.pnas.org/content/106/50/21173.full.pdf}
  {https://www.pnas.org/content/106/50/21173.full.pdf} \BibitemShut {NoStop}%
\bibitem [{\citenamefont {Eakins}\ \emph
  {et~al.}(2021{\natexlab{a}})\citenamefont {Eakins}, \citenamefont {Patel},
  \citenamefont {Kalra}, \citenamefont {Rezania}, \citenamefont {Shankar},\
  and\ \citenamefont {Tuszynski}}]{JT1}%
  \BibitemOpen
  \bibfield  {author} {\bibinfo {author} {\bibfnamefont {B.~B.}\ \bibnamefont
  {Eakins}}, \bibinfo {author} {\bibfnamefont {S.~D.}\ \bibnamefont {Patel}},
  \bibinfo {author} {\bibfnamefont {A.~P.}\ \bibnamefont {Kalra}}, \bibinfo
  {author} {\bibfnamefont {V.}~\bibnamefont {Rezania}}, \bibinfo {author}
  {\bibfnamefont {K.}~\bibnamefont {Shankar}}, \ and\ \bibinfo {author}
  {\bibfnamefont {J.~A.}\ \bibnamefont {Tuszynski}},\ }\href {\doibase
  10.3389/fmolb.2021.650757} {\bibfield  {journal} {\bibinfo  {journal}
  {Frontiers in Molecular Biosciences}\ }\textbf {\bibinfo {volume} {8}},\
  \bibinfo {pages} {150} (\bibinfo {year} {2021}{\natexlab{a}})}\BibitemShut
  {NoStop}%
\bibitem [{\citenamefont {Tuszynski}(2019)}]{Tuszynski2019}%
  \BibitemOpen
  \bibfield  {author} {\bibinfo {author} {\bibfnamefont {J.~A.}\ \bibnamefont
  {Tuszynski}},\ }\enquote {\bibinfo {title} {The bioelectric circuitry of the
  cell},}\ in\ \href {\doibase 10.1007/978-3-030-21293-3_11} {\emph {\bibinfo
  {booktitle} {Brain and Human Body Modeling: Computational Human Modeling at
  EMBC 2018}}},\ \bibinfo {editor} {edited by\ \bibinfo {editor} {\bibfnamefont
  {S.}~\bibnamefont {Makarov}}, \bibinfo {editor} {\bibfnamefont
  {M.}~\bibnamefont {Horner}}, \ and\ \bibinfo {editor} {\bibfnamefont
  {G.}~\bibnamefont {Noetscher}}}\ (\bibinfo  {publisher} {Springer
  International Publishing},\ \bibinfo {address} {Cham},\ \bibinfo {year}
  {2019})\ pp.\ \bibinfo {pages} {195--208}\BibitemShut {NoStop}%
\bibitem [{\citenamefont {Minoura}\ and\ \citenamefont
  {Muto}(2006)}]{Minoura2006}%
  \BibitemOpen
  \bibfield  {author} {\bibinfo {author} {\bibfnamefont {I.}~\bibnamefont
  {Minoura}}\ and\ \bibinfo {author} {\bibfnamefont {E.}~\bibnamefont {Muto}},\
  }\href {\doibase 10.1529/biophysj.105.071324} {\bibfield  {journal} {\bibinfo
   {journal} {Biophysical journal}\ }\textbf {\bibinfo {volume} {90}},\
  \bibinfo {pages} {3739} (\bibinfo {year} {2006})},\ \bibinfo {note}
  {16500962[pmid]}\BibitemShut {NoStop}%
\bibitem [{\citenamefont {R{\"u}hle}\ \emph {et~al.}(2011)\citenamefont
  {R{\"u}hle}, \citenamefont {Lukyanov}, \citenamefont {May}, \citenamefont
  {Schrader}, \citenamefont {Vehoff}, \citenamefont {Kirkpatrick},
  \citenamefont {Baumeier},\ and\ \citenamefont {Andrienko}}]{Ruhle2011}%
  \BibitemOpen
  \bibfield  {author} {\bibinfo {author} {\bibfnamefont {V.}~\bibnamefont
  {R{\"u}hle}}, \bibinfo {author} {\bibfnamefont {A.}~\bibnamefont {Lukyanov}},
  \bibinfo {author} {\bibfnamefont {F.}~\bibnamefont {May}}, \bibinfo {author}
  {\bibfnamefont {M.}~\bibnamefont {Schrader}}, \bibinfo {author}
  {\bibfnamefont {T.}~\bibnamefont {Vehoff}}, \bibinfo {author} {\bibfnamefont
  {J.}~\bibnamefont {Kirkpatrick}}, \bibinfo {author} {\bibfnamefont
  {B.}~\bibnamefont {Baumeier}}, \ and\ \bibinfo {author} {\bibfnamefont
  {D.}~\bibnamefont {Andrienko}},\ }\href {\doibase 10.1021/ct200388s}
  {\bibfield  {journal} {\bibinfo  {journal} {Journal of Chemical Theory and
  Computation}\ }\textbf {\bibinfo {volume} {7}},\ \bibinfo {pages} {3335}
  (\bibinfo {year} {2011})}\BibitemShut {NoStop}%
\bibitem [{\citenamefont {Kirkpatrick}\ \emph {et~al.}(2013)\citenamefont
  {Kirkpatrick}, \citenamefont {Lenzmann},\ and\ \citenamefont
  {Staffilani}}]{Kirkpatrick2013}%
  \BibitemOpen
  \bibfield  {author} {\bibinfo {author} {\bibfnamefont {K.}~\bibnamefont
  {Kirkpatrick}}, \bibinfo {author} {\bibfnamefont {E.}~\bibnamefont
  {Lenzmann}}, \ and\ \bibinfo {author} {\bibfnamefont {G.}~\bibnamefont
  {Staffilani}},\ }\href {\doibase 10.1007/s00220-012-1621-x} {\bibfield
  {journal} {\bibinfo  {journal} {Communications in Mathematical Physics}\
  }\textbf {\bibinfo {volume} {317}},\ \bibinfo {pages} {563} (\bibinfo {year}
  {2013})}\BibitemShut {NoStop}%
\bibitem [{\citenamefont {Eakins}\ \emph
  {et~al.}(2021{\natexlab{b}})\citenamefont {Eakins}, \citenamefont {Patel},
  \citenamefont {Kalra}, \citenamefont {Rezania}, \citenamefont {Shankar},\
  and\ \citenamefont {Tuszynski}}]{Eakins}%
  \BibitemOpen
  \bibfield  {author} {\bibinfo {author} {\bibfnamefont {B.~B.}\ \bibnamefont
  {Eakins}}, \bibinfo {author} {\bibfnamefont {S.~D.}\ \bibnamefont {Patel}},
  \bibinfo {author} {\bibfnamefont {A.~P.}\ \bibnamefont {Kalra}}, \bibinfo
  {author} {\bibfnamefont {V.}~\bibnamefont {Rezania}}, \bibinfo {author}
  {\bibfnamefont {K.}~\bibnamefont {Shankar}}, \ and\ \bibinfo {author}
  {\bibfnamefont {J.~A.}\ \bibnamefont {Tuszynski}},\ }\href {\doibase
  10.3389/fmolb.2021.650757} {\bibfield  {journal} {\bibinfo  {journal}
  {Frontiers in Molecular Biosciences}\ }\textbf {\bibinfo {volume} {8}},\
  \bibinfo {pages} {150} (\bibinfo {year} {2021}{\natexlab{b}})}\BibitemShut
  {NoStop}%
\bibitem [{\citenamefont {Schnyder}\ \emph {et~al.}(2009)\citenamefont
  {Schnyder}, \citenamefont {Ryu}, \citenamefont {Furusaki},\ and\
  \citenamefont {Ludwig}}]{Schnyder}%
  \BibitemOpen
  \bibfield  {author} {\bibinfo {author} {\bibfnamefont {A.~P.}\ \bibnamefont
  {Schnyder}}, \bibinfo {author} {\bibfnamefont {S.}~\bibnamefont {Ryu}},
  \bibinfo {author} {\bibfnamefont {A.}~\bibnamefont {Furusaki}}, \ and\
  \bibinfo {author} {\bibfnamefont {A.~W.~W.}\ \bibnamefont {Ludwig}},\ }\href
  {\doibase 10.1063/1.3149481} {\bibfield  {journal} {\bibinfo  {journal} {AIP
  Conference Proceedings}\ }\textbf {\bibinfo {volume} {1134}},\ \bibinfo
  {pages} {10} (\bibinfo {year} {2009})},\ \Eprint
  {http://arxiv.org/abs/https://aip.scitation.org/doi/pdf/10.1063/1.3149481}
  {https://aip.scitation.org/doi/pdf/10.1063/1.3149481} \BibitemShut {NoStop}%
\bibitem [{\citenamefont {Delplace}\ \emph {et~al.}(2011)\citenamefont
  {Delplace}, \citenamefont {Ullmo},\ and\ \citenamefont
  {Montambaux}}]{Delplace}%
  \BibitemOpen
  \bibfield  {author} {\bibinfo {author} {\bibfnamefont {P.}~\bibnamefont
  {Delplace}}, \bibinfo {author} {\bibfnamefont {D.}~\bibnamefont {Ullmo}}, \
  and\ \bibinfo {author} {\bibfnamefont {G.}~\bibnamefont {Montambaux}},\
  }\href {\doibase 10.1103/PhysRevB.84.195452} {\bibfield  {journal} {\bibinfo
  {journal} {Phys. Rev. B}\ }\textbf {\bibinfo {volume} {84}},\ \bibinfo
  {pages} {195452} (\bibinfo {year} {2011})}\BibitemShut {NoStop}%
\bibitem [{\citenamefont {Zhu}\ \emph {et~al.}(2019)\citenamefont {Zhu},
  \citenamefont {Prodan},\ and\ \citenamefont {Ahn}}]{Zhu}%
  \BibitemOpen
  \bibfield  {author} {\bibinfo {author} {\bibfnamefont {L.}~\bibnamefont
  {Zhu}}, \bibinfo {author} {\bibfnamefont {E.}~\bibnamefont {Prodan}}, \ and\
  \bibinfo {author} {\bibfnamefont {K.~H.}\ \bibnamefont {Ahn}},\ }\href
  {\doibase 10.1103/PhysRevB.99.041117} {\bibfield  {journal} {\bibinfo
  {journal} {Phys. Rev. B}\ }\textbf {\bibinfo {volume} {99}},\ \bibinfo
  {pages} {041117} (\bibinfo {year} {2019})}\BibitemShut {NoStop}%
\bibitem [{\citenamefont {Li}\ \emph {et~al.}(2017)\citenamefont {Li},
  \citenamefont {Lin}, \citenamefont {Zhang},\ and\ \citenamefont {Song}}]{Li}%
  \BibitemOpen
  \bibfield  {author} {\bibinfo {author} {\bibfnamefont {C.}~\bibnamefont
  {Li}}, \bibinfo {author} {\bibfnamefont {S.}~\bibnamefont {Lin}}, \bibinfo
  {author} {\bibfnamefont {G.}~\bibnamefont {Zhang}}, \ and\ \bibinfo {author}
  {\bibfnamefont {Z.}~\bibnamefont {Song}},\ }\href@noop {} {\bibfield
  {journal} {\bibinfo  {journal} {Physical Review B}\ }\textbf {\bibinfo
  {volume} {96}} (\bibinfo {year} {2017})}\BibitemShut {NoStop}%
\bibitem [{\citenamefont {H\"ugel}\ and\ \citenamefont
  {Paredes}(2014)}]{Hugel}%
  \BibitemOpen
  \bibfield  {author} {\bibinfo {author} {\bibfnamefont {D.}~\bibnamefont
  {H\"ugel}}\ and\ \bibinfo {author} {\bibfnamefont {B.}~\bibnamefont
  {Paredes}},\ }\href {\doibase 10.1103/PhysRevA.89.023619} {\bibfield
  {journal} {\bibinfo  {journal} {Phys. Rev. A}\ }\textbf {\bibinfo {volume}
  {89}},\ \bibinfo {pages} {023619} (\bibinfo {year} {2014})}\BibitemShut
  {NoStop}%
\bibitem [{\citenamefont {Padavi\ifmmode~\acute{c}\else \'{c}\fi{}}\ \emph
  {et~al.}(2018)\citenamefont {Padavi\ifmmode~\acute{c}\else \'{c}\fi{}},
  \citenamefont {Hegde}, \citenamefont {DeGottardi},\ and\ \citenamefont
  {Vishveshwara}}]{Padavic}%
  \BibitemOpen
  \bibfield  {author} {\bibinfo {author} {\bibfnamefont {K.}~\bibnamefont
  {Padavi\ifmmode~\acute{c}\else \'{c}\fi{}}}, \bibinfo {author} {\bibfnamefont
  {S.~S.}\ \bibnamefont {Hegde}}, \bibinfo {author} {\bibfnamefont
  {W.}~\bibnamefont {DeGottardi}}, \ and\ \bibinfo {author} {\bibfnamefont
  {S.}~\bibnamefont {Vishveshwara}},\ }\href {\doibase
  10.1103/PhysRevB.98.024205} {\bibfield  {journal} {\bibinfo  {journal} {Phys.
  Rev. B}\ }\textbf {\bibinfo {volume} {98}},\ \bibinfo {pages} {024205}
  (\bibinfo {year} {2018})}\BibitemShut {NoStop}%
\bibitem [{\citenamefont {Lieu}\ \emph {et~al.}(2020)\citenamefont {Lieu},
  \citenamefont {McGinley},\ and\ \citenamefont {Cooper}}]{Lieu}%
  \BibitemOpen
  \bibfield  {author} {\bibinfo {author} {\bibfnamefont {S.}~\bibnamefont
  {Lieu}}, \bibinfo {author} {\bibfnamefont {M.}~\bibnamefont {McGinley}}, \
  and\ \bibinfo {author} {\bibfnamefont {N.~R.}\ \bibnamefont {Cooper}},\
  }\href {\doibase 10.1103/PhysRevLett.124.040401} {\bibfield  {journal}
  {\bibinfo  {journal} {Phys. Rev. Lett.}\ }\textbf {\bibinfo {volume} {124}},\
  \bibinfo {pages} {040401} (\bibinfo {year} {2020})}\BibitemShut {NoStop}%
\bibitem [{\citenamefont {McGinley}\ and\ \citenamefont
  {Cooper}(2019)}]{McGinley}%
  \BibitemOpen
  \bibfield  {author} {\bibinfo {author} {\bibfnamefont {M.}~\bibnamefont
  {McGinley}}\ and\ \bibinfo {author} {\bibfnamefont {N.~R.}\ \bibnamefont
  {Cooper}},\ }\href {\doibase 10.1103/PhysRevResearch.1.033204} {\bibfield
  {journal} {\bibinfo  {journal} {Phys. Rev. Research}\ }\textbf {\bibinfo
  {volume} {1}},\ \bibinfo {pages} {033204} (\bibinfo {year}
  {2019})}\BibitemShut {NoStop}%
\bibitem [{\citenamefont {Campos~Venuti}\ \emph {et~al.}(2017)\citenamefont
  {Campos~Venuti}, \citenamefont {Ma}, \citenamefont {Saleur},\ and\
  \citenamefont {Haas}}]{Campos}%
  \BibitemOpen
  \bibfield  {author} {\bibinfo {author} {\bibfnamefont {L.}~\bibnamefont
  {Campos~Venuti}}, \bibinfo {author} {\bibfnamefont {Z.}~\bibnamefont {Ma}},
  \bibinfo {author} {\bibfnamefont {H.}~\bibnamefont {Saleur}}, \ and\ \bibinfo
  {author} {\bibfnamefont {S.}~\bibnamefont {Haas}},\ }\href {\doibase
  10.1103/PhysRevA.96.053858} {\bibfield  {journal} {\bibinfo  {journal} {Phys.
  Rev. A}\ }\textbf {\bibinfo {volume} {96}},\ \bibinfo {pages} {053858}
  (\bibinfo {year} {2017})}\BibitemShut {NoStop}%
\bibitem [{\citenamefont {P\'erez-Gonz\'alez}\ \emph
  {et~al.}(2019)\citenamefont {P\'erez-Gonz\'alez}, \citenamefont {Bello},
  \citenamefont {G\'omez-Le\'on},\ and\ \citenamefont {Platero}}]{NNN}%
  \BibitemOpen
  \bibfield  {author} {\bibinfo {author} {\bibfnamefont {B.}~\bibnamefont
  {P\'erez-Gonz\'alez}}, \bibinfo {author} {\bibfnamefont {M.}~\bibnamefont
  {Bello}}, \bibinfo {author} {\bibfnamefont {A.}~\bibnamefont
  {G\'omez-Le\'on}}, \ and\ \bibinfo {author} {\bibfnamefont {G.}~\bibnamefont
  {Platero}},\ }\href {\doibase 10.1103/PhysRevB.99.035146} {\bibfield
  {journal} {\bibinfo  {journal} {Phys. Rev. B}\ }\textbf {\bibinfo {volume}
  {99}},\ \bibinfo {pages} {035146} (\bibinfo {year} {2019})}\BibitemShut
  {NoStop}%
\bibitem [{\citenamefont {Vemu}\ \emph {et~al.}(2017)\citenamefont {Vemu},
  \citenamefont {Atherton}, \citenamefont {Spector}, \citenamefont {Moores},\
  and\ \citenamefont {Roll-Mecak}}]{Vemu2017-zh}%
  \BibitemOpen
  \bibfield  {author} {\bibinfo {author} {\bibfnamefont {A.}~\bibnamefont
  {Vemu}}, \bibinfo {author} {\bibfnamefont {J.}~\bibnamefont {Atherton}},
  \bibinfo {author} {\bibfnamefont {J.~O.}\ \bibnamefont {Spector}}, \bibinfo
  {author} {\bibfnamefont {C.~A.}\ \bibnamefont {Moores}}, \ and\ \bibinfo
  {author} {\bibfnamefont {A.}~\bibnamefont {Roll-Mecak}},\ }\href@noop {}
  {\bibfield  {journal} {\bibinfo  {journal} {Mol Biol Cell}\ }\textbf
  {\bibinfo {volume} {28}},\ \bibinfo {pages} {3564} (\bibinfo {year}
  {2017})}\BibitemShut {NoStop}%
\bibitem [{\citenamefont {Prodan}(2011)}]{EProdan}%
  \BibitemOpen
  \bibfield  {author} {\bibinfo {author} {\bibfnamefont {E.}~\bibnamefont
  {Prodan}},\ }\href {\doibase 10.1088/1751-8113/44/11/113001} {\bibfield
  {journal} {\bibinfo  {journal} {Journal of Physics A: Mathematical and
  Theoretical}\ }\textbf {\bibinfo {volume} {44}},\ \bibinfo {pages} {113001}
  (\bibinfo {year} {2011})}\BibitemShut {NoStop}%
\bibitem [{\citenamefont {Claes}\ and\ \citenamefont {Hughes}(2020)}]{Claes}%
  \BibitemOpen
  \bibfield  {author} {\bibinfo {author} {\bibfnamefont {J.}~\bibnamefont
  {Claes}}\ and\ \bibinfo {author} {\bibfnamefont {T.~L.}\ \bibnamefont
  {Hughes}},\ }\href {\doibase 10.1103/PhysRevB.101.224201} {\bibfield
  {journal} {\bibinfo  {journal} {Phys. Rev. B}\ }\textbf {\bibinfo {volume}
  {101}},\ \bibinfo {pages} {224201} (\bibinfo {year} {2020})}\BibitemShut
  {NoStop}%
\bibitem [{\citenamefont {Mondragon-Shem}\ \emph {et~al.}(2014)\citenamefont
  {Mondragon-Shem}, \citenamefont {Hughes}, \citenamefont {Song},\ and\
  \citenamefont {Prodan}}]{Hughes}%
  \BibitemOpen
  \bibfield  {author} {\bibinfo {author} {\bibfnamefont {I.}~\bibnamefont
  {Mondragon-Shem}}, \bibinfo {author} {\bibfnamefont {T.~L.}\ \bibnamefont
  {Hughes}}, \bibinfo {author} {\bibfnamefont {J.}~\bibnamefont {Song}}, \ and\
  \bibinfo {author} {\bibfnamefont {E.}~\bibnamefont {Prodan}},\ }\href
  {\doibase 10.1103/PhysRevLett.113.046802} {\bibfield  {journal} {\bibinfo
  {journal} {Phys. Rev. Lett.}\ }\textbf {\bibinfo {volume} {113}},\ \bibinfo
  {pages} {046802} (\bibinfo {year} {2014})}\BibitemShut {NoStop}%
\bibitem [{\citenamefont {Gadadhar}\ \emph {et~al.}(2017)\citenamefont
  {Gadadhar}, \citenamefont {Bodakuntla}, \citenamefont {Natarajan},\ and\
  \citenamefont {Janke}}]{Gadadhar}%
  \BibitemOpen
  \bibfield  {author} {\bibinfo {author} {\bibfnamefont {S.}~\bibnamefont
  {Gadadhar}}, \bibinfo {author} {\bibfnamefont {S.}~\bibnamefont
  {Bodakuntla}}, \bibinfo {author} {\bibfnamefont {K.}~\bibnamefont
  {Natarajan}}, \ and\ \bibinfo {author} {\bibfnamefont {C.}~\bibnamefont
  {Janke}},\ }\href {\doibase 10.1242/jcs.199471} {\bibfield  {journal}
  {\bibinfo  {journal} {Journal of Cell Science}\ }\textbf {\bibinfo {volume}
  {130}},\ \bibinfo {pages} {1347} (\bibinfo {year} {2017})},\ \Eprint
  {http://arxiv.org/abs/https://journals.biologists.com/jcs/article-pdf/130/8/1347/1946365/jcs199471.pdf}
  {https://journals.biologists.com/jcs/article-pdf/130/8/1347/1946365/jcs199471.pdf}
  \BibitemShut {NoStop}%
\end{thebibliography}%
\end{document}